# Band alignments of electronic energy structure in epitaxially grown β-Ga$_2$O$_3$ layers

D.A. Zatsepin[1,2], D.W. Boukhvalov[3,4], A.F. Zatsepin[1], Yu. A. Kuznetsova[1], D. Gogova[5,*],

V.Ya. Shur[6], A.A. Esin[6]

[1]*Institute of Physics and Technology, Ural Federal University, Mira Str. 19, 620002 Yekaterinburg, Russia*
[2]*M.N. Miheev Institute of Metal Physics of Ural Branch of Russian Academy of Sciences, 18 Kovalevskoj Str., 620990 Yekaterinburg, Russia*
[3]*Department of Chemistry, Hanyang University, 17 Haengdang-dong, Seongdong-gu, Seoul 133-791, Korea*
[4]*Theoretical Physics and Applied Mathematics Department, Ural Federal University, Mira Street 19, 620002 Yekaterinburg, Russia*
[5]*Central Lab of Solar Energy and New Energy Sources at the Bulg. Acad. Sci., Tzarigradsko shose72, 1784 Sofia, Bulgaria*
[6]*Institute of Natural Sciences, Ural Federal University, 51 Lenin Ave, 620000 Yekaterinburg, Russia*

**Abstract**

β-Ga$_2$O$_3$ epitaxial layers grown on β-Ga$_2$O$_3$ (100) and Al$_2$O$_3$ (0001) substrates were characerized by X-ray phtotelectron (XPS) and optical reflectance spectroscopies. The XPS electronic structure mapping combined with Density functional theory (DFT) calculations of densities of states allow to find and demonstrate different gallium-oxygen contributions to the Valence Band (VB) region of all investigated samples in the range of semi-core states at 15 – 25 eV and strong spectral suppression of the valence base-band (BB) area, which remains unchanged. An energy shift of the BB area from the top of the VB towards the higher binding energies has been established and it leads to an almost immutability of $E_g$ value in the β-Ga$_2$O$_3$ epilayers studied, despite of their dissimilar defectiveness. The results obtained well coincide with the DFT modeling of the final electronic structure and optical reflectance measurements.

[*]Corresponding author E-mail: dgogova@abv.bg

1. Introduction

The group of III-sesquioxides has been rediscovered and classified recently as a new class of wide band-gap semiconductors. During last decades polycrystalline highly doped (larger than $10^{21}$ cm$^{-3}$ – a degenerate state) indium and gallium oxide (In$_2$O$_3$:Sn(Ga)) thin films were employed as a transparent conductive material for transparent electrodes in "smart windows",[1,2] photovoltaics,[1] large-area flat panel displays,[3] etc. Nowadays, many efforts are focused on the development of single-crystalline III-sesquioxides with low defect densities and semiconducting behavior to be used as active layers in electronic and optoelectronic devices. The thermodynamically stable β-Ga$_2$O$_3$ phase is the most attractive representative of this class of materials due to its relatively large band-gap (~ 4.85 eV) promising applications for the area of short wavelength photonics and transparent electronics.[4] Moreover, it's high break-down field value of 8 MV/cm (exceeding that of Si, GaAs, SiC, III-nitrides and some technologically relevant oxides like ZnO) is very prospective for the high-power electronics as well. Also, Baliga's figure of merit for β-Ga$_2$O$_3$ is several times larger than that of 4H-SiC or GaN. Thus, the unique properties of this material, combined with the availability of simple and low-cost melt growth methods for large-scale production of bulk single crystals (float zone growth, edge-defined film-fed growth method and Czochralski growth method) in comparison to GaN substrates[5-7] actually have nominated β-Ga$_2$O$_3$ as a valuable candidate for the next-generation power electronics.

As a matter of principle, one might also expect that extremely wide band-gap semiconductors (where the band-gaps are essentially larger than 3 eV) may have another undisclosed potential for the challenging technologies since optoelectronic applications in the DUV region are emerging in biotechnology and nanotechnology. By doping this material with rare-earth elements or 3d transition-metal ions, Ga$_2$O$_3$ thin films have also demonstrate promising optical and photo-luminescent properties[8] as well as possible applications in electroluminescent devices (i.e., thin film electroluminescent displays).[9] In summary, since the β-Ga$_2$O$_3$ possesses one of the largest band gaps among all transparent conductive oxides, it has the potential to bring revolutionary performance to the next generation of optoelectronic devices, particularly to those operating in the deep UV region. So it is not a surprise that the electronic structure of β-Ga$_2$O$_3$ was the subject of hot discussions during last decade (see, i.e., Refs. 10-14). The cited papers yielded the challenging points concerning the characterization of the Valence Band (VB) electronic structure,[10] the core-like (semi-core) and Fermi level vicinity states,[11] standard DFT-based[12,13] and GGA-based[14] modeling of the symmetry type of oxygen vacant sites and possible displacements of Ga-atoms in a comparative analysis of the conventional and defective unit-cells. No doubt that these points are of great importance for the metal- and semimetal-embedding technologies employing β-Ga$_2$O$_3$ as a host-matrix, which initially has the wide-gap insulating origin from the very beginning. Despite the validity of scientific results present in cited above Refs., there are some missed challenging research items – nobody has performed the complete electronic structure mapping (including core-levels and VB Base-Band analysis) of β-Ga$_2$O$_3$, limiting their study with VB characterization only. Also, the high-quality XPS reference Ga-metal data is still absent even in the internationally accepted XPS Database probably because of the melting specificity of this metal and, hence, subsequent experimental difficulties of its XPS certification as well as the certification of Ga-containing compounds. All these make some obstacles while obtaining the XPS information about core-levels structure both for central metal-atom and for ligands regarding recognition the peculiarities of Ga–O chemical bond (i.e., dangling gallium-oxygen bonds, oxygen deficiency or oxygen excess) in the concrete material. It is obvious that the latter is not allowing to perform the technological treatment of these defects in a correct manner.[14]

In the current paper, we are presenting the results of complete electronic structure XPS-study (core-levels and valence band) as well as DFT calculations of the defects formation energies and densities of states of pure and defected "bulk" and "surface" β-Ga$_2$O$_3$ samples. The probable scenarios of electronic structure formation in epilayers of β-Ga$_2$O$_3$ and its alignments will be analyzed and discussed by comparison of the experimentally obtained data and theoretical models.

## 2. Experimental details and theoretical calculations

Metal-organic vapor phase epitaxy (MOVPE) has been selected as the method of choice in this study due to the high quality of the yielded crystalline material. This technology allows achieving the combination of relatively high growth rates and good homogeneity of the material deposited on large areas[15-19] which applies to the grade of electronic/optoelectronic device fabrication on industrial scale. The growth experiments were carried out in a low-pressure MOVPE top-fed reactor. Timethyl-gallium (Ga(CH$_3$)$_3$(TMGa)) as a source of gallium and water vapors as a source of oxygen were employed. First growth experiments have been performed on basal plane sapphire only to establish the growth window (temperature, base pressure, Ga/O ratio, etc.) for deposition of good quality β-Ga$_2$O$_3$ epitaxial layers and figured out the impact of each one of the growth parameters on the material crystallinity.[18,19] No any buffer layer was employed. However, to to obtain a damage-free surface with atomic steps the β-Ga$_2$O$_3$ (100) substrates were annealed in O$_2$-atmosphere at 950°C for 1 h. During the heating up and cooling down stages the substrates and epilayers were kept in O$_2$-containing ambient to prevent any accidental and undesirable surface decomposition. The reactor base pressure was fixed at 500 Pa. The Ar flow-rate through the water bubbler was varied in the range 300–750 sccm and the temperature of the bubbler of 50°C. The growth temperature of the β-Ga$_2$O$_3$ layers had been kept at 800°C and 825°C, and different molar fractions of the chemical regents (TMGa : O$_2$) were employed. The deposition time for this series of experiments was short to grow β-Ga$_2$O$_3$ layers with a thickness from 5 to 90 nm what was good enough for the XPS electronic structure qualifications. More details about the employed technological process could be found elsewhere.[18,19]

The following samples were obtained and named: Sample "1" (11-nm-thick β-Ga$_2$O$_3$ on β-Ga$_2$O$_3$ (100) substrate), Sample "2" (9-nm-thick β-Ga$_2$O$_3$ on Al$_2$O$_3$ (0001) substrate), Sample "3" (88-nm-thick β-Ga$_2$O$_3$ on Al$_2$O$_3$ (0001) substrate), and Sample "4" (13-nm-thick β-Ga$_2$O$_3$ on Al$_2$O$_3$ (0001) substrate). The lattice parameters of our homoepitaxially grown MOVPE β-Ga$_2$O$_3$ layers with a monoclinic structure have been detrmined by XRD as: $a$ = 1.23 nm, $b$ = 0.30 nm, $c$ = 0.58 nm and $\beta$ = 103.82 º which well coincide with those reported for β-$Ga_2O_3$ in Ref. 20.

Thermo Scientific™ *K*-Alpha+™ XPS spectrometer had been applied for the fast wide-scan chemical contamination analysis (survey XPS spectroscopy) and core-level and Valence Band (VB) structure XPS mapping. The selection of this spectrometer was based on the important advantages which are highly valuable in XPS spectroscopy for the precise electronic structure analysis – dual-beam neutralizer yielding the very-low-energy co-axial electrons (less than 10 eV) for excluding the charging of the insulating samples under analysis (GB Patent 2411763) and the 180º double-focusing hemispherical energy analyzer providing energy resolution not worth than 0.28 eV. The monochromatized Al *K*α X-ray source with 400 μm X-ray spot dia and not more than 70 W X-ray power load onto the sample in an oil-free vacuum at 5 ×$10^{-6}$ Pa pressure was employed. The XPS analyzer settings were: (*i*) 200 eV pass-energy window in a fast wide-scan mode (XPS survey spectra measurements); (*ii*) 50 eV pass-energy window for the core-level and VB mapping modes with a multi-point spectra acquisition (posterior XPS data summation and averaging). All measurements were made under CAE operation regime of the Thermo Scientific™ *K*-Alpha+™ XPS spectrometer. This means that the pass-energy window value of the hemispherical analyzer and its angular photoelectrons collecting energy resolution both remain constant throughout all-time binding energy scan, so the XPS data becomes more reliable and precise than in CRR mode where actual pass-energy is a variable. In fact, the CRR regime is not applicable for our measurements.

The results of XPS survey chemical contamination analysis of our samples will be interpreted on the cross-referencing basis of the most reliable and internationally accepted XPS databases,[21,22] Ref. 23 and Thermo Scientific "Thermo Avantage" Software version 5.951. We have selected the Sample "1" (11-nm-thick β-$Ga_2O_3$ homoepitaxially grown on β-$Ga_2O_3$ (100)-oriented substrate) as an XPS reference because in this case there is no lattice and thermal mismatch between the epilayer and the substrate and the only reason for structural deviations from the perfect monoclinic lattice will be intrinsic defects of the crystalline lattice caused by the growth conditions selected. In our humble opinion, this is a justified choice from materials science point of view.

The XPS Survey spectra shown in Fig. 1 have the distinct and clear sharp-peak structure which was well-recognized by "Thermo Avantage" Software at the elements qualification stage. The XPS databases crossed referencing[21-23] fully confirmed this qualification and allow to substantiate that no impurity XPS and Auger signals were detected except those that belong to the chemical composition of the current samples under study. The C 1s core-level signal, caused by the sprayed neutral carbon onto the surface of the samples for binding energy (BE) calibrations, is very weak but it is quite enough to perform precise BE-referencing. Finally, the performed analysis derives the suitability of samples "1 – 4" for the XPS electronic structure mapping (core-levels and valence bands).

Optical reflectance data of our samples were obtained using Perkin Elmer™ Lambda 35™ UV/VIS spectrophotometer. This spectrophotometer has true double-beam operations which provide the most possible stability in measurements using the comparance of obtained data with reference materials in real time, sealed and quartz-coated high-throughput optics for hi-brilliance and clear reflectance signal and fast-scanning real-time signal recording system in the operating range of wave-lengths 190 - 1100 nm (for deeper details in specifications, please, refer to the Manufacturer web-site).

The atomic structure and energetics of various configurations of defects in $Ga_2O_3$ nanoparticles were studied by the DFT method using the QUANTUM-ESPRESSO code[24] and the GGA–PBE[25] feasible for the modeling of impurities in oxides.[26] We used energy cutoffs of 25 Ry and 400 Ry for the plane-wave expansion of the wave functions and the charge density, respectively, and the 3×3×3 Monkhorst-Pack $k$-point grid for the Brillouin sampling[27] in the case of bulk and 3×3×2 in the case of (001) surface.

For the modeling of the bulk β-$Ga_2O_3$ the 2×2×2 supercell (80 atoms) is employed and for the modeling of the surface, we use the same supercell as a slab within the periodic boundary condition separated by 2 nm along the $c$ axis. The modeling of various defects has been made, such as single and double oxygen vacancy, gallium vacancy and gallium atom in the interstitial void. Formation energies of the surfaces are calculated by the formula:

$$E_{form} = (E_{surf} - E_{bulk})/n, \qquad (1)$$

where $E_{surf}$, $E_{bulk}$ are the total energies of the supercell of bulk and slab that contain the same number of atoms and $n$ is the number of atoms on the surface of the slab. Formation energy of the defects is calculated by the following formula:

$$E_{form} = (E_{host+defect} - (E_{host} \pm n\mu_{defect}))/n, \qquad (2)$$

where $E_{host}$, $E_{host+defect}$ represent the total energies of the pristine and defective bulk of surfaces; $\mu_{defect}$ – is the total energy of molecular oxygen in the ground (triplet) state of bulk gallium and $n$ – denotes the number of defect atoms. The sign inside the parentheses is negative for the case of vacancies formation and positive for the case if interstitial impurities.

### 3. Results and discussions

*3.1 Primary regions core-level XPS*

We will start the electronic structure XPS mapping of our samples from the analysis of the primary XPS region – Ga 2p core-levels – notwithstanding the fact that these electronic states are quite far away from $E_F$ vicinity (see the actual BE-values of XPS Ga 2p in the XPS Survey shown in Fig. 1 and, separately recorded with higher energy resolution, in Fig. 2).

Gallium metal of high purity (99.99999 %) was employed as an XPS $Ga^0$ reference (Merck, the former Sigma-Aldrich, USA) and combined {Ga-metal + oxidized Ga}[22] XPS Ga $2p_{3/2-1/2}$ external standard[22] in order to detect precisely the BE-shift between Ga 2p core-levels in the oxidized $Ga^{3+}$ and $Ga^0$ charge states. Recall, that up to present still no any other charge states of Ga were found in its oxide form except 3+. As one can clearly see from Fig. 2, the BE-shift among Ga $2p_{3/2-1/2}$ core-levels in Ga-metal and Ga-oxide XPS external standards is about 2.1 eV, thus allowing to separate these charge states of Ga in a quite easy way opposite to that for Zn-metal and ZnO where this BE-shift is less that 0.3 eV.[22] The Ga$2p_{3/2-1/2}$ spin-orbitally separated spectral components have visually seen essentially asymmetric line-shapes for metallic gallium whereas being oxidized these asymmetric line-shapes become symmetric (see Fig. 2, compare red and grey spectra). The actual BE-values of Ga$2p_{3/2-1/2}$ spectral components are given below in Table 1.

This well-known XPS difference feature for metals and oxides allows to affirm that in a combined {Ga-metal + oxidized Ga} XPS standard[22] we have the signs of rather slightly acidified Ga-Ox and stoichiometric $Ga_2O_3$ mixture than pure Ga-metal + $Ga_2O_3$ one: (*i*) both double Ga $2p_{3/2}$ and double Ga $2p_{1/2}$ components exhibits visually symmetrical line-shapes which are characteristic for compounds; (*ii*) the BE-difference between very high-pure Ga-metal XPS 2p spectrum and that for reference standard reported in Ref. 22 is visually recognized as well (compare red and grey spectra in Fig. 2). The different intensity ratio of these components in Ga $2p_{3/2}$ and Ga $2p_{1/2}$ spin-orbitally separated bands means that the $Ga_2O_3$ content is dominating in compliance with Ga-Ox so this XPS reference standard with the high probability is mostly of oxide origin and, apparently, have only negligible content of pure Ga-metal because of its high acidification ability at normal ambient.

The same double-band Ga $2p_{3/2}$ and double-band Ga $2p_{1/2}$ structure of XPS Ga 2p core-level has Samples "3" and "4" (see Fig. 2, upper part) for the same discussed above reason as conditionally {Ga-metal + oxidized Ga} XPS standard, but, opposite to it, one can see the nearly equal contributions of Ga-Ox and $Ga_2O_3$ phases to the structure of XPS Ga 2p core-levels in these samples. At the same time Samples "1" and "2" exhibiting single-band Ga $2p_{3/2}$ and single-band Ga $2p_{1/2}$ spin-orbital components of XPS Ga 2p, with that having strongly dissimilar BE locations – symmetrical line-shapes of Ga 2p with nearly ideal coincidence with "$Ga_2O_3$ part" of conditionally {Ga-metal + oxidized Ga} XPS standard for Sample "1" and as well symmetrical line-shapes of Ga 2p for Sample "2", but with the intermediate BE-positions of the Ga $2p_{3/2}$ and Ga $2p_{1/2}$ between Ga-metal and $Ga_2O_3$. Good agreement for the $Ga_2O_3$ part of the spectrum of conditionally {Ga-metal + oxidized Ga} XPS standard and Sample "1" is not a surprise because as it had been shown above this sample is of β-$Ga_2O_3$ origin, what was just spectrally proved one more time. Regarding the analysed case of matter, the intermediate BE positions of Ga $2p_{3/2}$ and Ga $2p_{1/2}$ components between Ga-metal and $Ga_2O_3$ possibly means that the defective oxygen phase is present in Sample "2", but the type of Ga–Ox bonding is closer to the stoichiometric β-$Ga_2O_3$. One might speculate about existance of other $Ga_2O_3$ phases in our layers, however, it is well-known they are not stable at the growth temperatures employed. So the most reasonable explanation seems to be another type of Ga ligand surroundings and imperfections. This supposition will be checked in the onward DFT modeling of the atomic and electronic structure of our layers.

XPS O 1s core-level spectra of Samples "2 – 4" are shown in Fig. 3. Comparing these spectra with the β-$Ga_2O_3$ XPS external standard (or Sample "1"), it is seen the strong dissimilarity among their symmetry, FWHM, and positions of their sub-bands. While the O 1s core-level spectrum of β-$Ga_2O_3$ (Sample "1") has completely symmetrical line-shape, the others are deviating from it essentially. The BE-position of O 1s symmetrical mono-band for β-$Ga_2O_3$ was reported earlier as 531.15 eV,[28] what is well coinciding with our 531.09 eV XPS data. The perfect coincidence regarding line-shape symmetry and BE-position of the O 1s core-level of Sample "1" is easily explained by the technological aspect of homoepitaxial growth on β-$Ga_2O_3$ (100) substrate. Recall, that the lattice and thermal mismatch in homoepitaxy equals to 0 %, so proving one more time our choice of Sample "1" and its spectral XPS mapping as an XPS external standard.

For the Samples "2 – 4" one can see the strong transformation of the O 1s core-level profiles, which is starting in the O 1s spectrum of Sample "2" as an appearance of visually detected asymmetry using additional 532 eV sub-band (see Fig. 3, black spectrum). This 532 eV XPS sub-band was well studied up to present and it is directly linked with oxygen sublattice defects, what had been reasonably proved in our previous XPS-and-DFT findings. The defective origin of this sub-band in the case of Sample "2" as well easily agrees with the fact that an attempt to grow β-

Ga$_2$O$_3$ on Al$_2$O$_3$ (0001) substrate is made and the finally yielded material has the structure which contains oxygen defects. Formally, there is a chance to speculate about the essentially different O 1s line-shape profile of Sample "2", which is eliminating the general spectral pattern of Samples "1-4". Because of the very thin epilayer (9-nm-thick β-Ga$_2$O$_3$) and heteroepitaxial growth on highly mismatched Al$_2$O$_3$ (0001) substrate, there is an in-plane tilt, between grains rotated 120° to each other, due to the difference in the crystal symmetry of the substrate (hexagonal) and the epilayer (monoclinic),[19] one might link the metal-oxygen bonding in this sample, on the basis of similarity of this bond to Al$_2$O$_3$, whose O 1s spectrum has nearly the same asymmetrical line-shape.[23] Moreover, one of the aluminum oxide polymorphs – alumina, has the same BE position as our Sample "2", namely 531.1 eV.[21,22] Nevertheless, the FWHM's are essentially dissimilar (∼ 1.6 – 1.7 eV for most Al$_2$O$_3$ polymorphs versus 1.3 eV for Sample "2") due to different asymmetry based on the different 532 eV band contributions and O 1s tails are essentially incompatible.[23] Therefore, we cannot assume that O 1s core-level of Sample "2" is of the same origin as in alumina.

The situation with O 1s core-levels of Samples "3 – 4" is another: here no any spectral signatures of defects were detected, but dual bands XPS core-level spectra structure is arising. One can see, the first XPS maxima coincides well with that of β-Ga$_2$O$_3$ XPS external standard (Sample "1") and have the same BE-location, and the second bands are shifted towards higher BE's, being totally outside from the common BE-range of O 1s for metal oxides[22,23] – 533 eV. This means that the structure of Samples "3 – 4" has two types of gallium-oxygen chemical bonding: the first one is similar to that of β-Ga$_2$O$_3$ and the other is of a different origin. We suppose that the second type of Ga–O appears because of the Al$_2$O$_3$ (0001) substrate influence on the final atomic structure of Samples "3 – 4" (lattice and thermal mismatch between substrate and epilayer), and, if our suppositions are valid, we will observe these peculiarities in the O 2s valence bands region of these samples. One more time, the BE position of the second O 1s band located at 532.8 eV of Samples "3-4" is not matching with that of Al$_2$O$_3$ polymorphs as well as the FWHMs are strongly dissimilar[22,23] with all the following consequences.

*3.2 Valence band and valence base-band XPS*

X-ray photoelectron Valence Band (VB) spectra of our MOVPE gallium oxide layers and metallic gallium are shown in Fig. 4. These spectra are characterized by the essentially weak densities of states (DOS) at the $E_F$ vicinity and, as one can see, relatively weak transformations of the Base-band area (BBA) in the range from 1 eV up to 10 eV. At the same time an essential transformations of the XPS VB line-shapes occur in the range of semi-core or core-like states (see the 15 – 25 eV area, Fig. 4). Generally speaking, the mentioned character of VB transformations is not so obvious for the conventional metal oxides where essential transformations of the VB-structure usually occur exactly in the range of $E_F$ vicinity DOSes, so the core-like states area of the VB of our samples will be analyzed primarily.

The most intensive set of narrow single-line XPS peaks of the samples and the Ga-metal at 18.5 eV belongs to the Ga $3d_{5/2 – 3/2}$ electronic states, exhibiting exactly small spin-orbital separation $\Delta = 0.46$ eV only,[22] and thus, unfortunately, not allowing to resolve these spin-orbital components in an easy way of XPS measurements. The Ga 3d electronic states origin of the 18.5 eV XPS band as well supported by our XPS measurements of Ga-metal of 99.99999 % purity taken as an XPS $Ga^0$ reference (see Fig. 4, red spectrum). Comparing this part of the VB among Ga-metal and Sample "1" (reference of β-Ga$_2$O$_3$), one might see that an additional band arises at 21.6 eV in the Sample's "1" appropriate spectrum. Recall, that this band is in the usual range of O 2s electronic states. In fact, it is not a surprise because of the oxide origin of Sample "1", so that's why it satisfactory coincides with that reported in XPS Databases.[21,22] This 21.6 eV band as well is present in the spectrum of Sample "2" and have the very close intensity to this band in Sample "1" (reference of β-Ga$_2$O$_3$), thus additionally confirming its origin. The Samples "3 – 4" are not visually exhibiting this 21.6 eV O 2s band, however, a new high-intensity 20.2 eV peak arises (see Fig. 4, blue and olive spectra). Wherein we have to note the contributing to the 20.2 eV peak character of the 21.6 eV band, and there is no doubt about it. According to the XPS Databases,[21,22] the 20.2 eV peak was interpreted previously as Ga 3d states from native gallium oxide. In our humble opinion, this interpretation is not completely correct if only because of the overlapping with 21.6 eV O 2s. Additionally, oxygen atoms as ligands are located in a distorted cubic closest packing arrangement yielding the Ga–O bond distances of 1.83 and 2.00 Å in native gallium oxide,[29] so allowing to identify the 20.2 eV peak rather as Ga 3d + O 2s (II) states with a small additive of O 2s (I) states than only as Ga 3d states. In favor of our point is the XPS O 1s core-level analysis (see Fig. 3) where exactly for Samples "3 – 4" the dissimilar O 1s states location at different binding energies has been detected. Thus, we might conclude that dissimilar types of Ga–O bondings were established experimentally in our XPS findings for Samples "3 – 4". Recall, these samples have an epitaxial relationship with the hexagonal sapphire substrate, however, there is an in-plan tilt[18] which might be a justified reason for different Ga–O bond distances, reported previously for the native type of gallium oxidation.[25] From photoionization cross-sections relation for the given Al $K\alpha$ excitation[30] σ (Ga 3d): σ (O 2s) = 1.41 : 0.19 the dominating character of Ga 3d becomes clear, so the range of binding energies from 15 eV to 25 eV is of Ga 3d majority with a small admixture of O 2s states.

The XPS mapping of Valence Base-Band Area (BBA) allows to state about rather a band-tail contribution of Ga 4p states in such a way that the origin of the BBA is determined exactly with O 2p electronic states (see Fig. 5) despite the σ (Ga 4p): σ (O 2p) = 1.54: 0.52 relation for the given Al $K\alpha$ excitation.[30] The BB-width remains almost nearly of the same value (～ 7.1 eV) and is only shifting from the top of the VB towards the higher binding energies, exhibiting the transformations of the majority of O 2p states because of defects in the oxygen sublattice. This finding well coincides with the reported above XPS core-level Ga 2p and O 1s analysis and will be checked in the onward DFT-based modeling of the electronic structure.

*3.3 DFT electronic structure modeling*

Calculations of the formation energies of various defects in β-$Ga_2O_3$ and the influence of these imperfections onto the electronic structure were the first steps of our DFT-modeling. The formation energy cost of various defects in β-Ga$_2$O$_3$ is rather high (see Table 2). Thus, one can conclude that β-Ga$_2$O$_3$ is rather robust and the concentration of the defects might be negligible. To perform the modeling of the film structure more realistic, we have performed the calculations of the void in the bulk of β-Ga$_2$O$_3$ by removing single gallium and four nearest oxygen atoms. This type of defect has a formation energy of 3.45 eV/atom, and this value allows to conclude that similar type of defects are more probable than the point defects discussed above. Formation of the void provides an energy shift of the VB-edge, of about 0.3 eV up (see Fig. 6a), and this is in qualitative agreement with the experimental XPS VB spectra shown in Fig. 5. On the other hand, the structure of lower energy levels, laying in -12 ~ -20 eV, remains nearly unchanged, thus, contradicting with the experimental results (Fig. 4). Therefore, we can valuably conclude that the changes in the electronic structure discussed above cannot be initiated by defects in the bulk β-Ga$_2$O$_3$.

The next step of our modeling was the evaluation of the contribution of the surface states to the electronic structure. We have performed the calculations of formation energy employing three types of surfaces – (100), (010) and (001) and found out that the lowest one has the minimal value of formation energy (0.62 eV/atom). Therefore, the (001) surface might be discussed as a feasible model of naturally formed grain boundaries in β-Ga$_2$O$_3$ films. Usually, the formation of the surface provides a visible broadening of the valence band (Fig. 6b). In contrast to the case of the void in the bulk β-Ga$_2$O$_3$, the surface states also lead to broadening of the Ga 3d and O 2p bands what is a good in agreement with the experimental XPS data illustrated in Fig. 4. Formation of various defects in the (001) surface is rather probable because of the small energy cost of this process. The presence of these defects does not significantly affect the electronic structure of the (001) surface at all (see Fig. 6b). These results correspond well to the similarities in the experimentally XPS-mapped electronic structure of the samples under investigation (Fig. 5), which has been synthesized under different conditions and, thus, without any doubts, are containing different configurations of defects. Therefore, we can conclude the formation of grain boundaries in β-Ga$_2$O$_3$ layers provides an appearance of the surface-like features in the final electronic structure. The most important point of electronic and optical structural transformations in the electronic structure is the shift of the upper limit of the valence band of about 0.2 eV up to the Fermi level.

*3.4 Determination of the optical band gap*

At the final stage of the current study, an estimation of the forbidden band-gap of β-Ga$_2$O$_3$ epilayers was made employing an optical reflectance technique. Figure 7 visualizes the experimental reflectance data of our layers. Analytical data processing of spectral dependencies in the region of the fundamental absorption edge has been performed using the Kumar model:[31]

$$D(h\nu) = \ln[(R_{max} - R_{min})/(R(h\nu) - R_{min})], \qquad (3)$$

where $D(h\nu)$ means the optical density; $R(h\nu)$ denotes experimental data obtained by means of optical measurements; $R_{max}$ and $R_{min}$ – maximum and minimum values of optical reflectance within the analyzing spectral range (4.7-5.3 eV in our current case).

The value of the band-gap might be determined with the help of the following equation:

$$D(h\nu) \cdot h\nu = A(h\nu - E_g)^n, \qquad (4)$$

where $A$ is a constant; $E_g$ means the forbidden band-gap; $n$ is the index determining the type of interband transitions (equals to 1/2 for direct allowed transitions and 3/2 – для for forbidden ones; in the case of indirect allowed and forbidden equals to 2 and 3, respectively[32]). One have to note, that the most perfect approximation of experimental data might achieved using $n = 1/2$, therefore indicating that the direct allowed transitions occur.

Figure 8 displays the spectral dependencies of the optical density constructed as functions in corresponding coordinates. Red lines in the graphs denote the approximation of the linear portion of the experimental data. The intersection of the line portions with the axis of abscissa corresponds to the value of the forbidden band-gap for direct transitions. The values obtained are presented in Table 3. As one can see, the band-gap values of the Samples "2", "3" and "4" are quite close being in the range of 5.23 – 5.30 eV. The results obtained satisfactory coincide with the XPS data and theoretically modeled electronic structure of $Ga_2O_3$ epilayers. At the same time, the optical data for Sample "1" demonstrates the typically reported in the literature value of 4.86 eV.

## 4. Conclusions

The XPS analysis of the electronic structure of β-$Ga_2O_3$ epilayers demonstrates that Samples "1" – "2" (11-nm-thick β-$Ga_2O_3$ on β-$Ga_2O_3$ (100) substrate and 9-nm-thick $Ga_2O_3$ on $Al_2O_3$ (0001) substrate, respectively) and "3" – "4" (88-nm-thick $Ga_2O_3$ on $Al_2O_3$ (0001) and 13-nm-thick $Ga_2O_3$ on $Al_2O_3$ (0001), respectively) have dissimilar types of gallium-oxygen chemical bondings what is the reason for the different gallium and oxygen contributions to the VB region in the range of semi-core states located at 15 – 25 eV and spectral suppression of the valence base-band (BB) area. At the same time, the BB-width remains unchanged, only shifting from the top of the VB towards the higher binding energies and exhibiting the transformations of the majority of O 2p states because of the defects in the oxygen sublattice. These experimentally established peculiarities of the electronic bands alignments are well supported by the DFT calculations as well as by the optical reflectance data. The latter allows to conclude about the almost immutability of $E_g$ value of the studied β-$Ga_2O_3$ epilayers despite of their dissimilar defectiveness. The analysis performed allows to conclude that the growth method and substrates employed yield materials with very similar electronic structure and optical properties, what is advantageous in terms of applications.

**Acknowledgements**

The XPS electronic structure study was supported by the Ministry of Education and Science of the Russian Federation (Government Task No. 3.1485.2017/4.6). The MOVPE $Ga_2O_3$ samples were grown at the IKZ-Berlin in the frame of the SAW-2012-IKZ-2 Project. We thank Z. Galazka for providing a Czochralski grown substrate for the MOVPE growth of sample 1.

**References:**

1. C. G. Granqvist, Handbook of Inorganic Electrochromic Materials, Ed.: C.G. Granqvist, 1st Edition, Elsevier (1995) 1.
2. D. Gogova, A. Iossifova, T. Ivanova, Z. Dimitrova and K. Gesheva, *J. Cryst. Growth*, 1999, **198-199**, 1230-1234.
3. U. Betz, M. Kharrazi Olsson, J. Marthy, M. F. Escolá and F. Atamny, *Surf. Coat. Technol.*, 2006, **200**, 5751.
4. J. Wager, *Science*, 2003, **300**, 1245-1246.
5. M. Hermann, D. Gogova, D. Siche, M. Schmidbauer, B. Monemar, M. Stutzmann and M. Eickhoff, *J. Cryst. Growth*, 2006, **293 (2)**, 462-468.
6. D. Gogova, P. P. Petrov, M. Buegler, M. R. Wagner, C. Nenstiel, G. Callsen, M. Schmidbauer, R. Kucharski, M. Zajac, R. Dwilinski, M. R. Phillips, A. Hoffmann and R. Fornari, *J. Appl. Phys.*, 2013, **113**, 203513.
7. D. Gogova, H. Larsson, A. Kasic, G. R. Yazdi, I. Ivanov, R. Yakimova, B. Monemar, E. Aujol, E. Frayssinet, J. -P. Faurie, B. Beaumont and P. Gibart, *Jap. J. Appl. Phys.*, 2005, **44 (3)**, 1181-1185.
8. A. Shionoya and W. M. Yen, Phosphor Handbook, Eds.: S. Shionoya, W. M. Yen and H. Yamamoto, 1st Edition, CRC Press, Boca Raton, FL (1998) 1.
9. T. Miyata, T. Nakatani and T. Minami, *J. Lumin.*, 2000, **87-89**, 1183-1185.
10. M. Mohamed, C. Janowitz, I. Unger, R. Manzke, Z. Galazka, R. Uecker, R. Fornari, J. R. Weber, J. B. Varley and C. G. Van de Walle, *Appl. Phys. Lett.*, 2010, **97**, 211903.
11. C. Janowitz, V. Scherer, M. Mohamed, A. Krapf, H. Dwelk, R. Manzke, Z. Galazka, R. Uecker, K. Irmscher, R. Fornari, M. Michling, D. Schmeiβer, J. R. Weber, J. B. Varley and C. G. Van de Walle, *New J. Phys.*, 2011, **13**, 085014.
12. C. Tang, J. Sun, N. Lin, Z. Jia, W. Mu, X. Tao and X. Zhao, *RSC Adv.*, 2016, **6**, 78322.
13. Y. Jinliang and Q. Chong, *J. Semicond.*, 2016, **37**, 042002-1.
14. J. B. Varley, J. R. Weber, A. Janotti and C. G. Van de Walle, *Appl. Phys. Lett.*, 2010, **97**, 142106.
15. G. B. Stringfellow, Organometallic Vapor-Phase Epitaxy: Theory and Practice, Academic Press, 2nd Edition (1999) 1.
16. D. Gogova, H. Larsson, R. Yakimova, Z. Zolnai, I. Ivanov and B. Monemar, *Phys. Status Solidi* (a), 2003, **200 (1)**, 13-17.
17. S. Rafique, L. Han, M. J. Tadjer, J. A. Freitas Jr., N. A. Mahadik and H. Zhao, *Appl. Phys. Lett.*, 2016, **108**, 182105.
18. D. Gogova, M. Schmidbauer and A. Kwasniewski, *Cryst. Eng. Comm.*, 2015, **17**, 6744-6752.
19. D. Gogova, G. Wagner, M. Baldini, M. Schmidbauer, K. Irmscher, R. Schewski, Z. Galazka, M. Albrecht and R. Fornari, *J. Cryst. Growth*, 2014, **401**, 665-669.

20. J. Åhman, G. Svensson and J. Albertsson, *Acta Cryst.*, 1996, **C52**, 1336-1338.
21. NIST Government XPS Database (Web-version), rev. 4.1, http://srdata.nist.gov/xps/ (called 2017-08-15).
22. Thermo Scientific XPS: Knowledge Base (Web-version), © 2013-2016, http://xpssimplified.com/knowledgebase.php (called 2017-08-15).
23. B. V. Crist, The PDF Handbooks of Monochromatic Spectra (Web-version, Vol.2), Ed.: XPS International LLC, USA, © 2005, 956 p., http:// www.xpsdata.com (called 2017-08-15).
24. P. Giannozzi, S. Baroni, N. Bonini, M. Calandra, R. Car, C. Cavazzoni, D. Ceresoli, G. L. Chiarotti, M. Cococcioni, I. Dabo, A. Dal Corso, S. de Gironcoli, S. Fabris, G. Fratesi, R. Gebauer, U. Gerstmann, C. Gougoussis, A. Kokalj, L. Michele, L. Martin-Samos, N. Marzari, F. Mauri, R. Mazzarello, S. Paolini, A. Pasquarello, L. Paulatto, C. Sbraccia, S. Scandolo, G. Sclauzero, A.P. Seitsonen, A. Smogunov, P. Umari and R. M. Wentzcovitch, *J. Phys.: Condens. Matter*, 21 (2009) 395502.
25. J. P. Perdew, K. Burke and M. Ernzerhof, *Phys. Rev. Lett.*, 1996, **77**, 3865.
26. D. A. Zatsepin, D. W. Boukhvalov, N. V. Gavrilov, A. F. Zatsepin, V. Ya. Shur, A. A. Esin, S. S. Kim and E. Z. Kurmaev, *Appl. Surf. Sci.*, 2017, **400**, 110-117.
27. H. J. Monkhorst and J. D. Pack, *Phys. Rev. B,* 1976, **13**, 5188.
28. V. D. Wheeler, D. I. Shanin, M. J. Tadjer and C. R. E. Eddy Jr., *ECS J. Solid State Sci. Technol.,* 2017, **6**, Q-3052.
29. R. B. King, Encyclopedia of Inorganic Chemistry, vol. 3, 2[nd] Edition, J. Wiley and Sons (2005) ISBN 978-0-470-86078-6, p.1256.
30. Atomic Calculation of Photoionization Cross-Sections and Asymmetry Parameters (Web-version), Elettra, Italy; https://vuo.elettra.eu/services/elements/WebElements.html (called 2017-09-08).
31. V. Kumar, S. Kr. Sharma, T. P. Sharma and V. Singh, *Optical Materials,* 1999, **12**, 115.
32. J. Tauc, Amorphous and Liquid Semiconductors, Plenum, New York (1974) ISBN: 0306307774, p.159-220.

**Figure captions:**

**Figure 1.** XPS Survey spectra of the Samples "1 - 4". Sample 1 is taken as an XPS external standard (see the reason explanations in the text above).

**Figure 2.** XPS Ga $2p_{3/2–1/2}$ core-levels of Samples "1 - 4" compared with that for Ga-metal and combined {Ga-metal + oxidized Ga}[22] XPS Ga $2p_{3/2–1/2}$ external standards.

**Figure 3.** XPS O 1s core-level spectra of Samples "2 – 4" comparing with that for Sample "1", taken as β-$Ga_2O_3$ XPS external standard. Note, that the intensities of all spectra were normalized with referencing the intensity of β-$Ga_2O_3$.

**Figure 4.** XPS Valence Band spectra of the Samples "2 – 4" comparing with that of the Sample "1", taken as the β-$Ga_2O_3$ XPS external standard, and the high-pure Ga-metal.

**Figure 5.** Zoomed XPS Valence Base-Band Area of Samples "2 – 4" compared with that of the Sample "1", taken as the β-$Ga_2O_3$ XPS external standard, and the high-pure Ga-metal.

**Figure 6.** Densities of States (DOSes): (a) for the bulk and (b) for the (001) surface of β-$Ga_2O_3$ with the most energetically favorable defects and without.

**Figure 7.** Optical reflectance spectra of the MOVPE β-$Ga_2O_3$ epilayers.

**Figure 8.** Spectral dependencies of the optical density of the MOVPE β-$Ga_2O_3$ layers built in the coordinates for allowed interband transitions.

**Table 1.** XPS parameters of Ga2p$_{3/2-1/2}$ core-level spectra for the samples and appropriate XPS external standards. BE-values for double-band spectral components are given in parentheses if any.

| Sample or XPS standard (reference) | Core-level component Binding Energy position (eV) | | Spin-orbital separation Δ (eV) |
|---|---|---|---|
| | Ga 2p$_{3/2}$ | Ga 2p$_{1/2}$ | |
| Ga-metal (reference) | 1116.5 | 1143.3 | 26.8 |
| Ga-metal+Ga$_2$O$_3$ (reference) | 1116.6 (1118.2) | 1143.4 (1145.0) | 26.8 (26.8) |
| Sample "1" (β-Ga$_2$O$_3$ reference) | 1118.3 | 1145.1 | 26.8 |
| Sample "2" | 1117.4 | 1144.2 | 26.8 |
| Sample "3" | 1116.9 (1118.4) | 1143.7 (1145.2) | 26.8 (26.8) |
| Sample "4" | 1116.7 (1118.5) | 1143.5 (1145.3) | 26.8 (26.8) |

**Table 2.** Formation energies per defect atom (in eV) for various configurations of defects in bulk and (001) surface of β-Ga$_2$O$_3$. The most probable configurations are marked in bold font.

| Configuration | Bulk | Surface |
|---|---|---|
| pure | --- | **+0.62** |
| vO | +3.83 | **-0.09** |
| 2vO | +3.90 | +0.80 |
| vGa | +9.14 | +4.24 |
| iGa | +11.87 | **-0.25** |

**Table 3.** Band-gap values of the MOVPE β-Ga$_2$O$_3$ epitaxial layers

| Sample No: | (eV) |
|---|---|
| 1 | 4.86 |
| 2 | 5.24 |
| 3 | 5.23 |
| 4 | 5.30 |

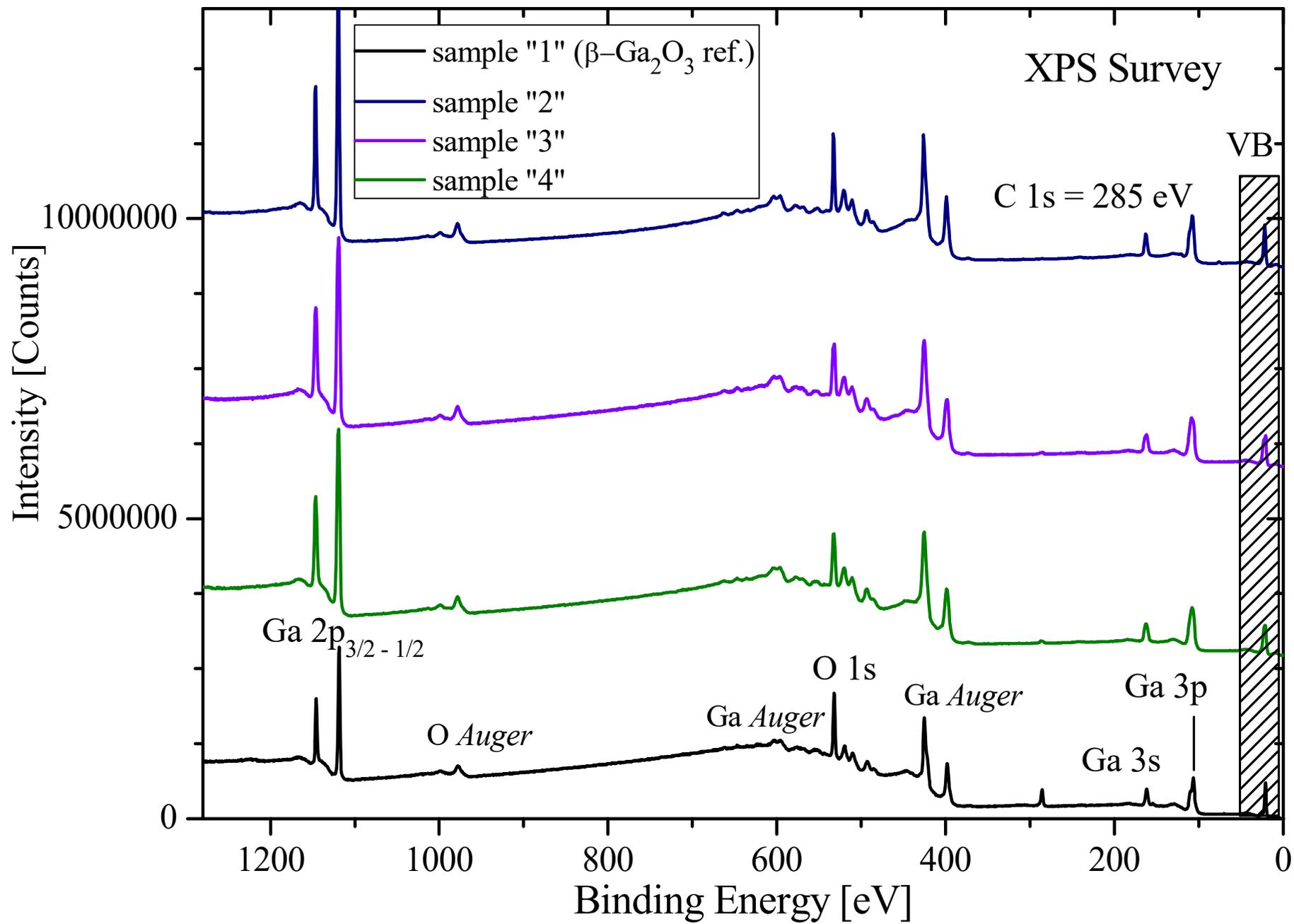

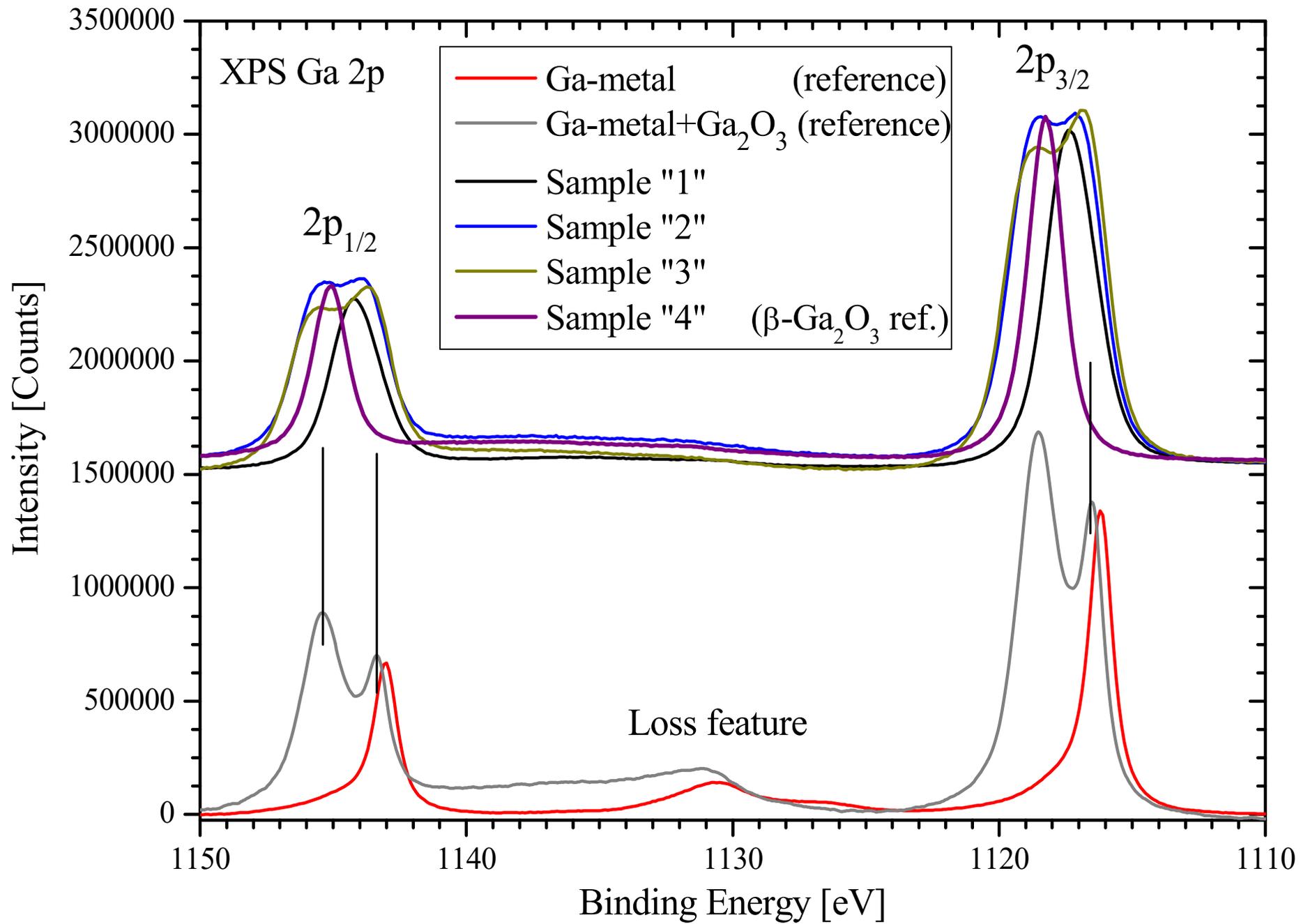

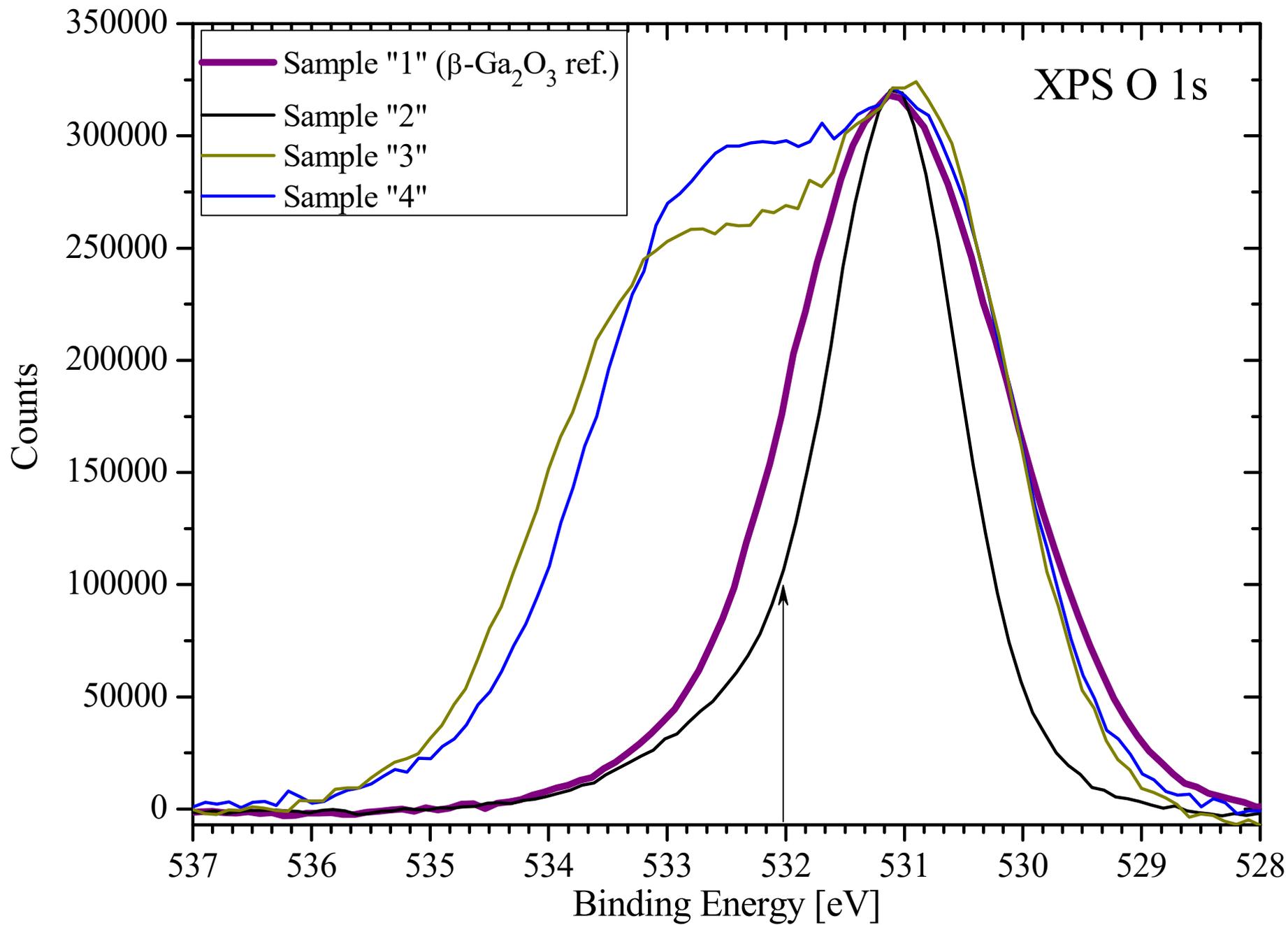

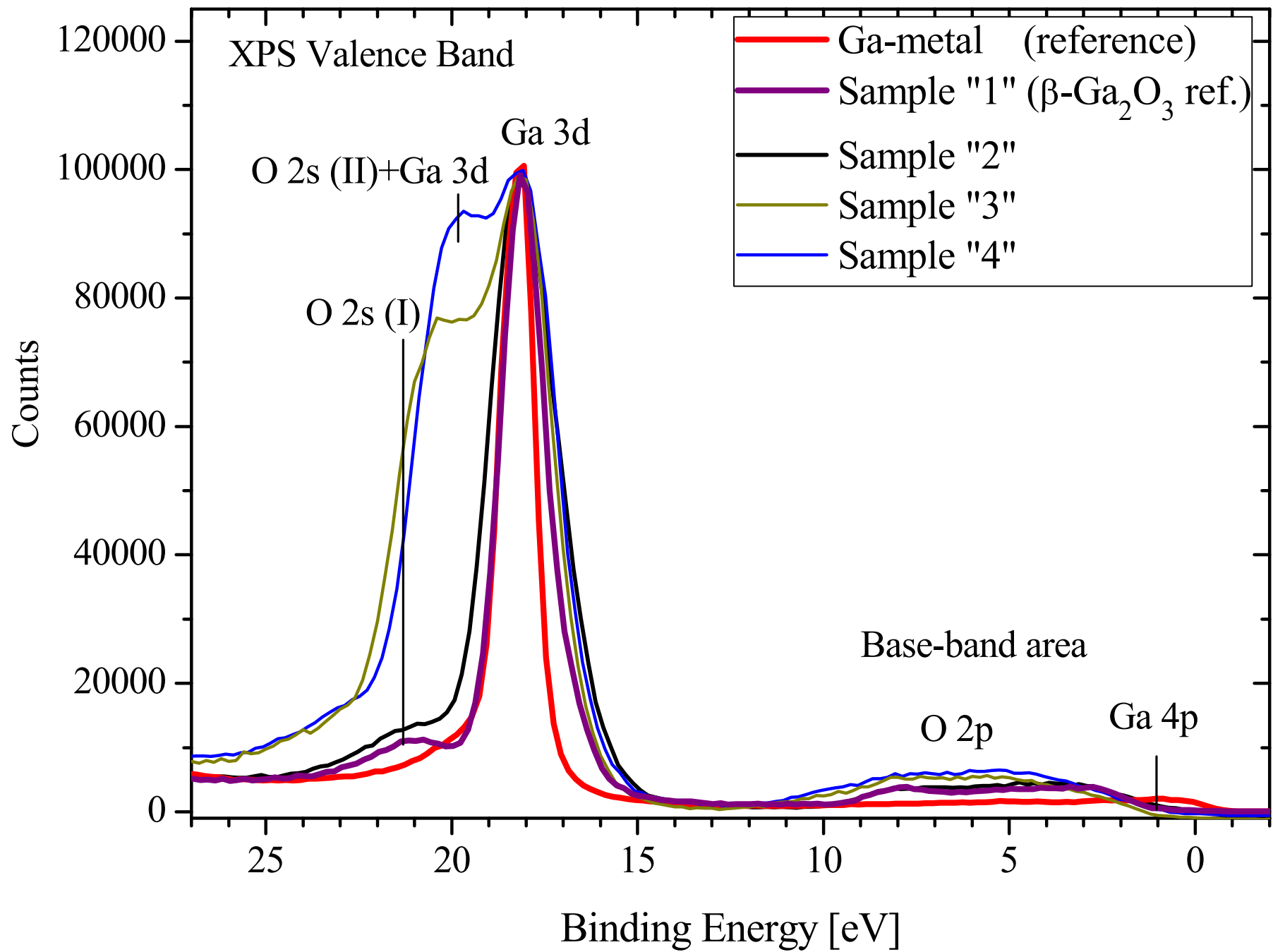

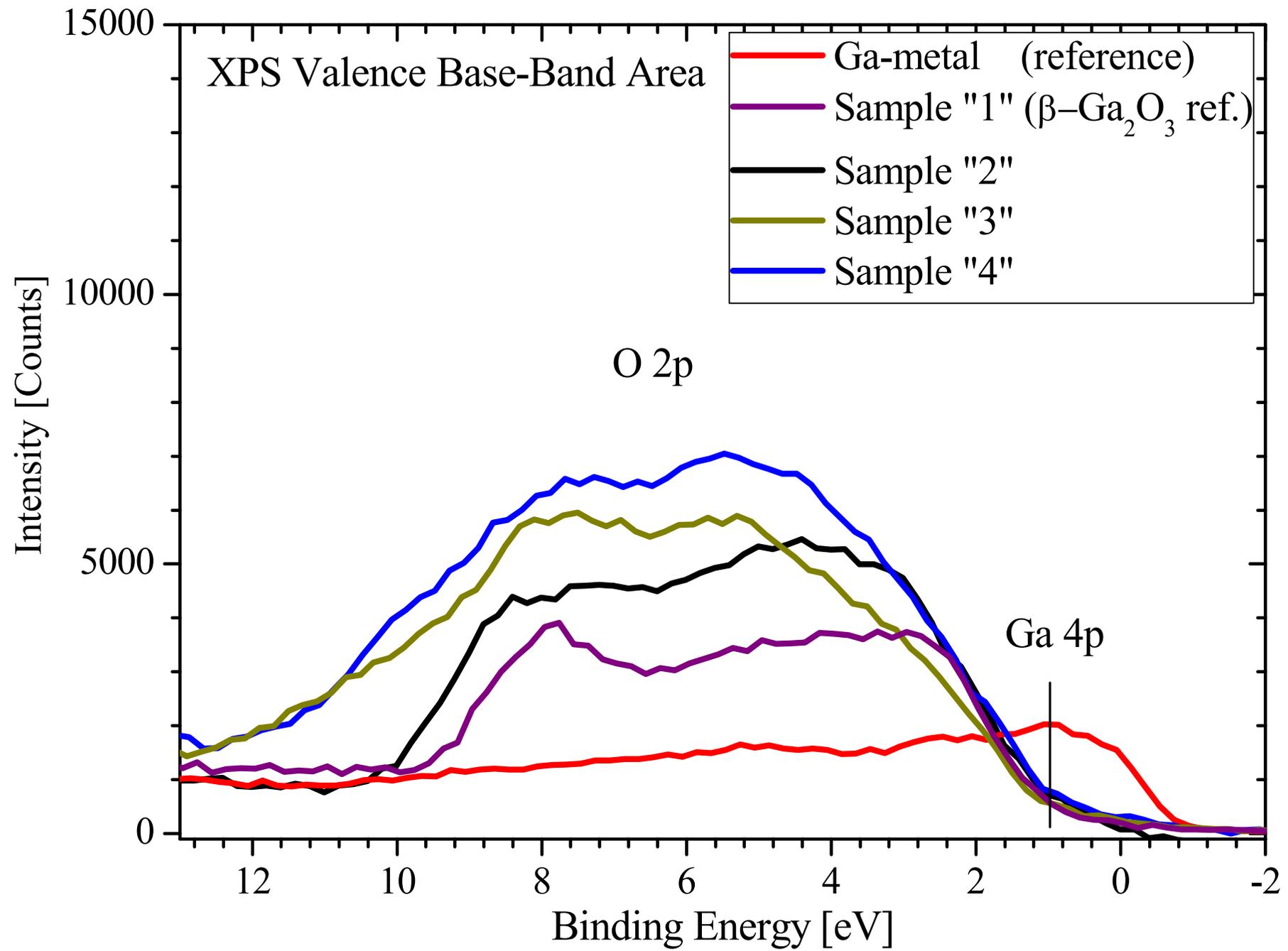

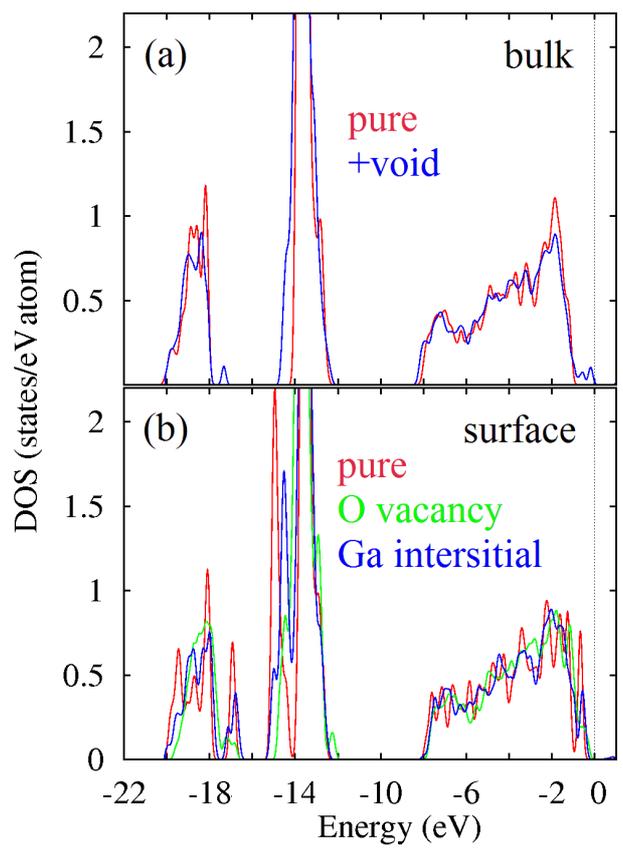

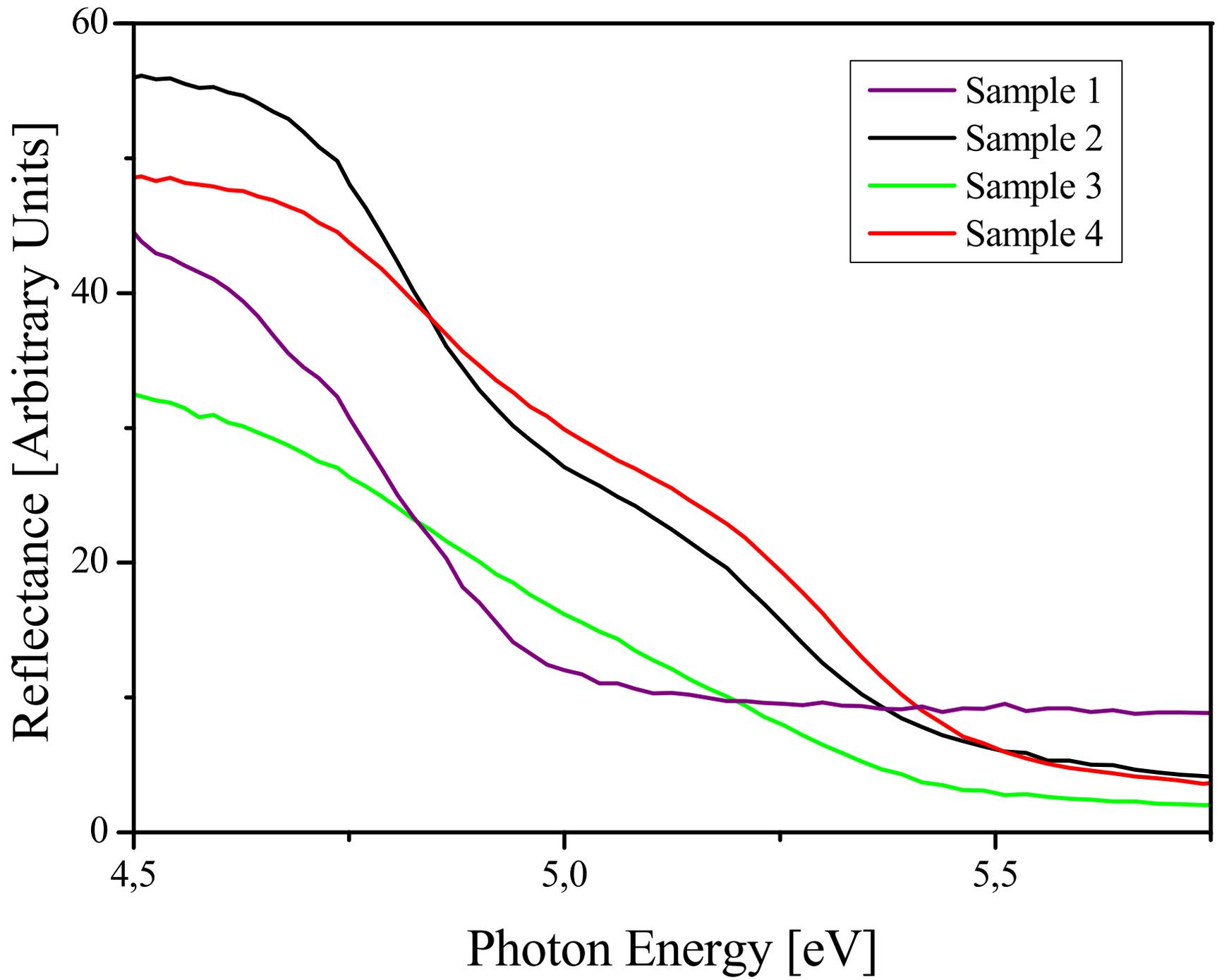

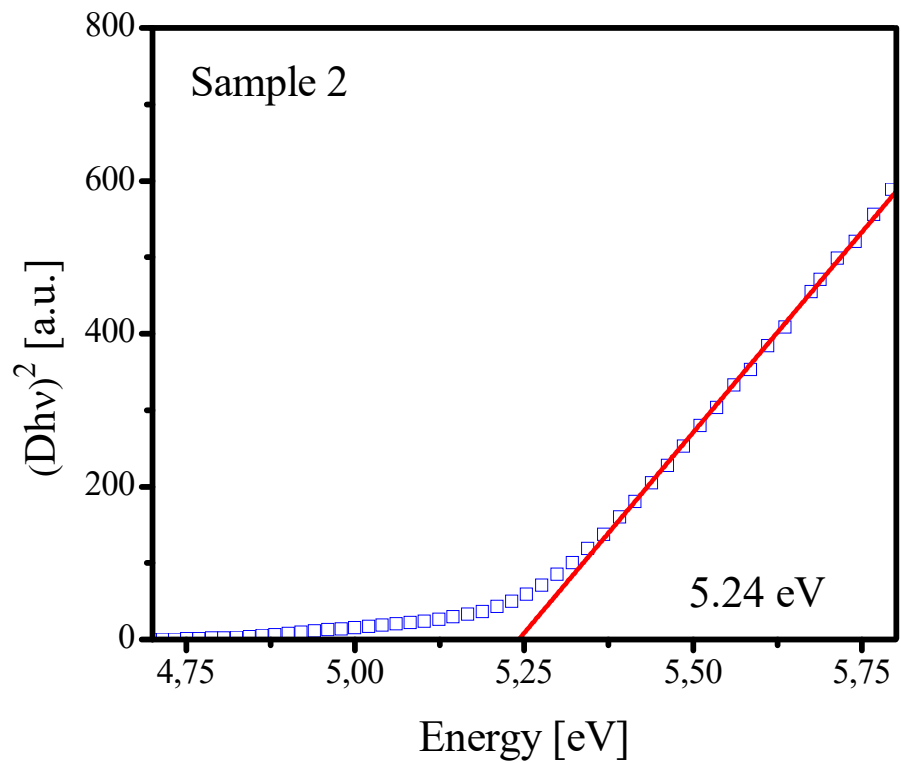
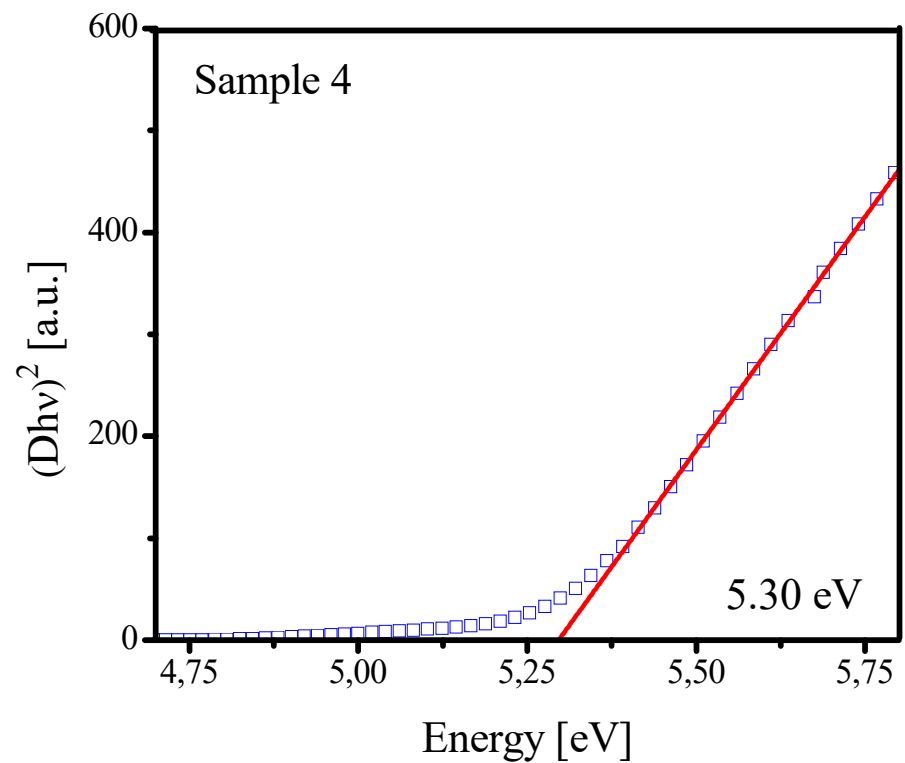
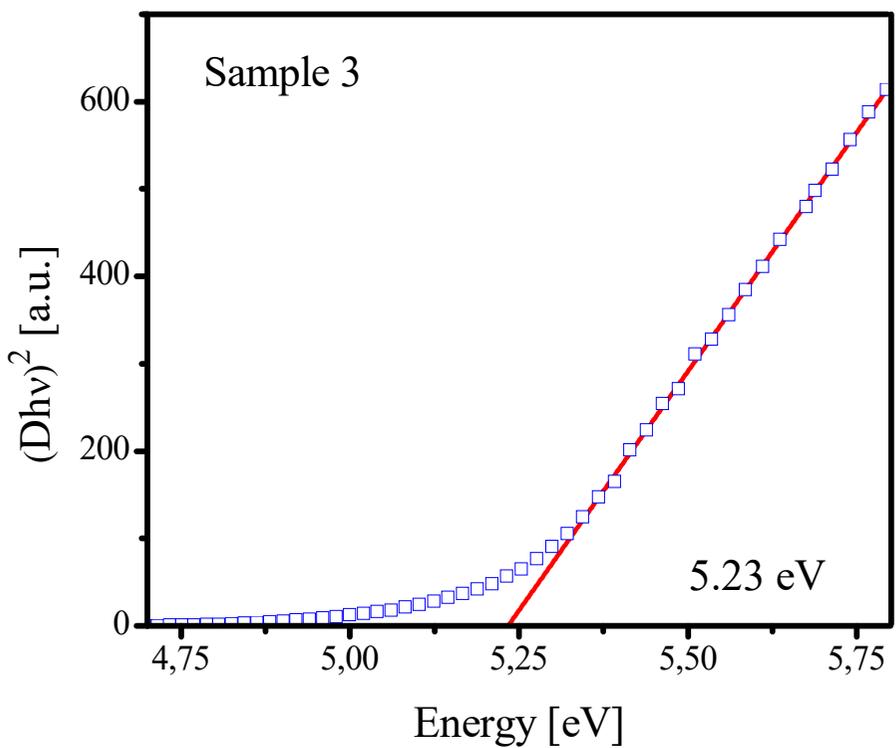
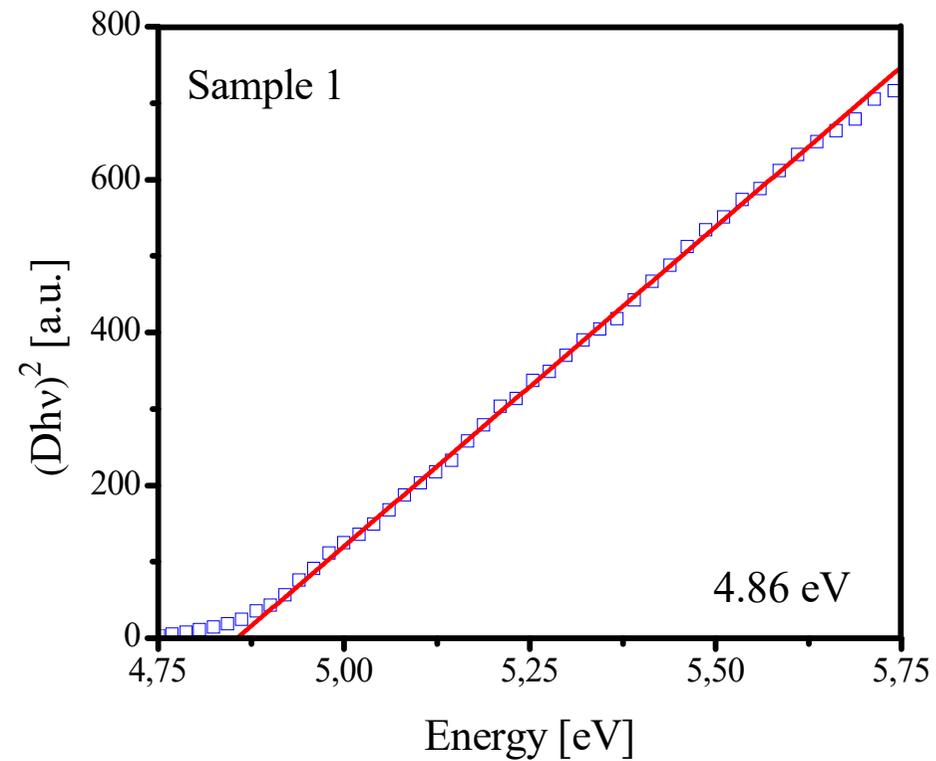